%% file: sample-sigconf.tex
\documentclass[sigconf]{acmart}

\usepackage{amsmath} 
\DeclareMathOperator*{\argmax}{argmax} 

\usepackage{algorithm}
\usepackage{algorithmic}
\usepackage{blindtext, graphicx}
\usepackage{amsfonts}
\usepackage{booktabs}
\usepackage{siunitx}
\usepackage{multirow}
\usepackage{threeparttable}
\usepackage{subfig}
\AtBeginDocument{%
  \providecommand\BibTeX{{%
    \normalfont B\kern-0.5em{\scshape i\kern-0.25em b}\kern-0.8em\TeX}}}

\acmDOI{10.1145/3321707.3321711}
\acmISBN{978-1-4503-6111-8/19/07}
\acmConference[GECCO '19]{the Genetic and Evolutionary Computation Conference 2019}{July 13--17, 2019}{Prague, Czech Republic}
\acmYear{2019}
\copyrightyear{2019}

\begin{document}

\title{Efficient Personalized Community Detection via Genetic Evolution}

\author{Zheng Gao}
\affiliation{\institution{Indiana University Bloomington}}
\affiliation{IN, United States}
\email{gao27@indiana.edu}

\author{Chun Guo}
\affiliation{\institution{Pandora Media LLC}}
\affiliation{CA, United States}
\email{cguo@pandora.com}

\author{Xiaozhong Liu}
\affiliation{\institution{Indiana University Bloomington}}
\affiliation{IN, United States}
\email{liu237@indiana.edu}


\begin{abstract}
Personalized community detection aims to generate communities associated with user need on graphs, which benefits many downstream tasks such as node recommendation and link prediction for users, etc. It is of great importance but lack of enough attention in previous studies which are on topics of user-independent, semi-supervised, or top-K user-centric community detection. Meanwhile, most of their models are time consuming due to the complex graph structure.  Different from these topics, personalized community detection requires to provide higher-resolution partition on nodes that are more relevant to user need while coarser manner partition on the remaining less relevant nodes. In this paper, to solve this task in an efficient way, we propose a genetic model including an offline and an online step. In the offline step, the user-independent community structure is encoded as a binary tree. And subsequently an online genetic pruning step is applied to partition the tree into communities. To accelerate the speed, we also deploy a distributed version of our model to run under parallel environment. Extensive experiments on multiple datasets show that our model outperforms the state-of-arts with significantly reduced running time.

\end{abstract}




\maketitle
\input{content/introRevised}
\input{content/reviewRevised}
\input{content/methodRevised}
\input{content/experimentRevised}
\input{content/conclusionRevised} 

\bibliographystyle{ACM-Reference-Format}
\bibliography{acmart} 

\end{document}

%% file: content/introRevised.tex
\section{Introduction}

Community detection is an important topic in graph mining. By learning node community labels on the graph, we are able to detect node hidden attributes as well as explore the closeness between nodes \cite{fortunato2010community,zhang2016others}. Conventional methods are mostly user-independent to detect communities solely relying on graph topological structure \cite{xia2017internal}, generate semi-supervised communities with node constraints, or select top-K sub graphs as user-centric communities. These approaches are no longer enough to satisfy users with a pursuit of personalization, which makes involving user need into community detection to become an inevitable task. Specifically, from a user-centric viewpoint, the ideal communities should provide a high-resolution partition in areas of the graph relevant to the user need and a coarse manner partition on the remaining areas so as to best depict user need (\textit{we also call it ``query'' in the rest of this paper}) in concentrated areas while fuzz irrelevant areas. 

\begin{figure}
	\center
	\includegraphics[width=1\columnwidth]{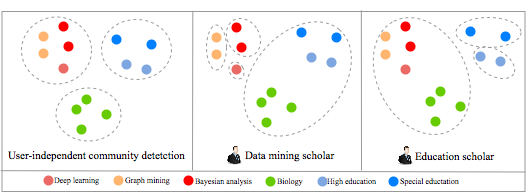} 
	\caption{An example of personalized community detection on a scholarly graph} 
	\label{fig:example}
\end{figure}

For instance, in Figure \ref{fig:example}, two different scholars in \textit{education} and \textit{data mining} domains may consume the same scholarly graph differently because they may need more detailed community exploration in their own domains while generalized community information in other irrelevant domains (e.g., the \textit{data mining} scholar needs more detailed communities such as \textit{Deep Learning}, \textit{Graph Mining}, and \textit{Bayesian Analysis}.  While an \textit{education} scholar may need to generalize those communities as \textit{Computer Science} or just \textit{Science}). 

As aforementioned, current investigations are still with limited scope. First, user-independent approaches solely consider graph topological structure without user need. For instance, \cite{yang2016modularity}  proposes a novel nonlinear reconstruction method by adopting deep neural networks to generate communities with the maximum modularity. \cite{chamberlain2018real} calculates the Jacaard similarity of neighborhood graphs to decide whether two nodes belong to the same community. Second, semi-supervised approaches detect communities restricted by pre-selected seed nodes. As different user needs refer to different seeds, each individual user requires a separate process to run the whole model completely to get personalized communities, which is inapplicable in real cases. \cite{ma2019community} introduces a joint approach to decompose the matrices associated with multi-layer networks and prior information into a community matrix and multiple coefficient matrices. \cite{gujral2018smacd} defines a non-negativity and a latent sparsity constraint to guide community detection. Third, sub-graph selection approaches only generate communities from the partial graph instead of the whole one. \cite{gupta2014top} designs two index infrastructures including a topology index and a metapath weight index to exploit top ranked subgraphs. Similarly, \cite{zou2007top} indexes the graph and uses cluster coefficient as the criteria for sub-graph selection. 

To detect personalized communities on the whole graph, in this paper, we propose a \textbf{g}enetic \textbf{P}ersonalized \textbf{C}ommunity \textbf{D}etection (gPCD) model with an offline and an online step. Specifically, in the offline step, we convert the user-independent graph community to a binary community tree which is encoded with binary code. Subsequently, a deep learning method is utilized to learn low-dimensional embedding representations for both user need and nodes on the binary community tree. In the online step, we propose a genetic tree-pruning approach on the tree to detect personalized communities by maximizing user need and minimizing user searching cost simultaneously. The whole genetic approach runs in an iterative manner to simulate an evolutionary process and generate a number of partition candidates which are regarded as ``chromosomes'' in each genetic generation. Through the selection, cross-over and mutation process, successive chromosomes are bred as better personalized community partitions to meet with user need.

The contribution of this study is threefold. 
\begin{itemize}
	\item We address a novel personalized community problem and propose a model to generate different-resolution communities associated with user need.   
	\item Our model contains an offline and an online step. The offline step takes charge of most calculation to enable an efficient online step: The construction of binary community tree has a time complexity of $O(\rVert V\lVert^{2})$ in the worst case where $\rVert V\lVert$ denotes the number of vertices in the graph; Representation learning on both binary community tree and user need has the same time complexity as Node2vec \cite{grover2016node2vec}. The online genetic pruning step running under the parallel environment achieves $O(\frac{2^dKP}{M})$ time complexity where $d$ denotes the depth of the tree, $K$ denotes the community number, $P$ denotes the initialized population size in the genetic approach and $M$ denotes the number of Mappers/ Reducers in Hadoop Distributed File System (HDFS). 
	\item We evaluate our model on a scholarly graph and a music graph. In our model, the offline step is separately calculated and keeps unchanged once constructed, while the online step guides the personalized community detection. Hence we only compare the online step results with baselines' performance. Extensive experiments shows our model outperforms in terms of both accuracy and efficiency.
\end{itemize}	

%% file: content/reviewRevised.tex
\section{Literature Review}

The problem of exploring community structure in graphs has long been a central research topic in network science \cite{fortunato2016community,chakraborty2017metrics,liu2016comparing}. From user-centric viewpoint, existing community detection methods can be divided into three categories: user-independent models, semi-supervised models and top-K community selection models.

\textbf{User-independent Community Detection:}  Models belonging to this category aims to generate communities solely relying on graph structure without considering any auxiliary information. As ``modularity'' is a classic metric to evaluate community quality \cite{girvan2002community}, there are a bunch of works which try to generate community partitions by maximizing graph modularity \cite{newman2006modularity,blondel2008fast,de2012novel}. Dynamic models can handle higher order structures with hierarchical communities \cite{benson2016higher}. Random walk dynamics are by far the most exploited track in community detection. For example, Infomap \cite{rosvall2008maps} detects communities by minimizing the description length for random walk paths. An extended Multilevel Infomap models reveals hierarchical community structure in a complex graph \cite{rosvall2011multilevel}. Recent works start to leverage deep learning methods for community detection . DeepWalk \cite{perozzi2014deepwalk} learns node embeddings based on random walks and Kmeans is subsequently applied on node embeddings to detect communities. Node2vec \cite{grover2016node2vec} and edge2vec \cite{gao2018edge2vec} are both extended models of DeepWalk which design a biased random walk to learn node embeddings better representing graph structure. \cite{yang2016modularity} aims to maximize the graph modularity with a deep learning framework. \cite{abdelbary2014utilizing} also uses neural networks to generate content-based communities for online social networks. Some other methods derived from statistical models are also able to detect communities. However, they are usually too time consuming to apply on large scale graphs \cite{karrer2011stochastic,law2017deep,abbe2018community}.

\textbf{Semi-supervised Community Detection:} In this track, community detection models are restricted by a pre-defined constraint \cite{gao2017personalized}. 
\cite{gujral2018smacd} using multi-aspect information of heterogeneous graph and a small ratio of node constraints to detect both non-overlapping and overlapping communities. \cite{li2018enhanced} detects communities by actively selecting a small amount of links as side information to sharpen the boundaries between communities and compact the connections within communities. \cite{ma2018semi} proposes a new community measurement metric and a spectral model to detect communities. \cite{ganji2018lagrangian} introduces a new constrained community detection model based on Lagrangian multipliers to incorporate and fully satisfy the node labels and pairwise constraints. \cite{wang2018unified} designs a unified non-negative matrix factorization framework simultaneously for community detection and semantic matching by integrating both semi-supervised information and node content. \cite{ma2010semi} discusses the equivalence of the objective functions of the symmetric non-negative matrix factorization (SNMF) and the maximum optimization of modularity density first, and derives the community detection model from the equivalence.

\textbf{Top-K Community Selection:} Some prior works also exploit on social graphs to find top-K groups of nodes that are relevant to a user query. \cite{lappas2009finding} addresses the problem of forming a team of skilled individuals based on a given task, while minimizing
the communication cost among the members of the team. In order to find the top K most relevant subgraphs given a user query, \cite{gupta2014top} introduces an offline approach generating two index structures for the network: a topology index, and a graph maximum
metapath weight index. An online novel top-K approach is subsequently applied to exploit these indexes for answering queries. To solve the same problem, \cite{zou2007top} designs a balanced tree (G-Tree) to index the large graph first and then proposes the rank matching (RM) algorithm to locate the top-k matches of query Q by pruning on the balanced tree. \cite{zhu2012finding} develops a new graph distance
measure using the maximum common subgraph (MCS), which is more accurate than the feature based measures, to find top-K similar graphs for a user query and offers an optimization approach to accelerate running speed.

%% file: content/methodRevised.tex
\section{genetic Personalized Community Detection}

\begin{figure}  
 \includegraphics[width=1.0\columnwidth]{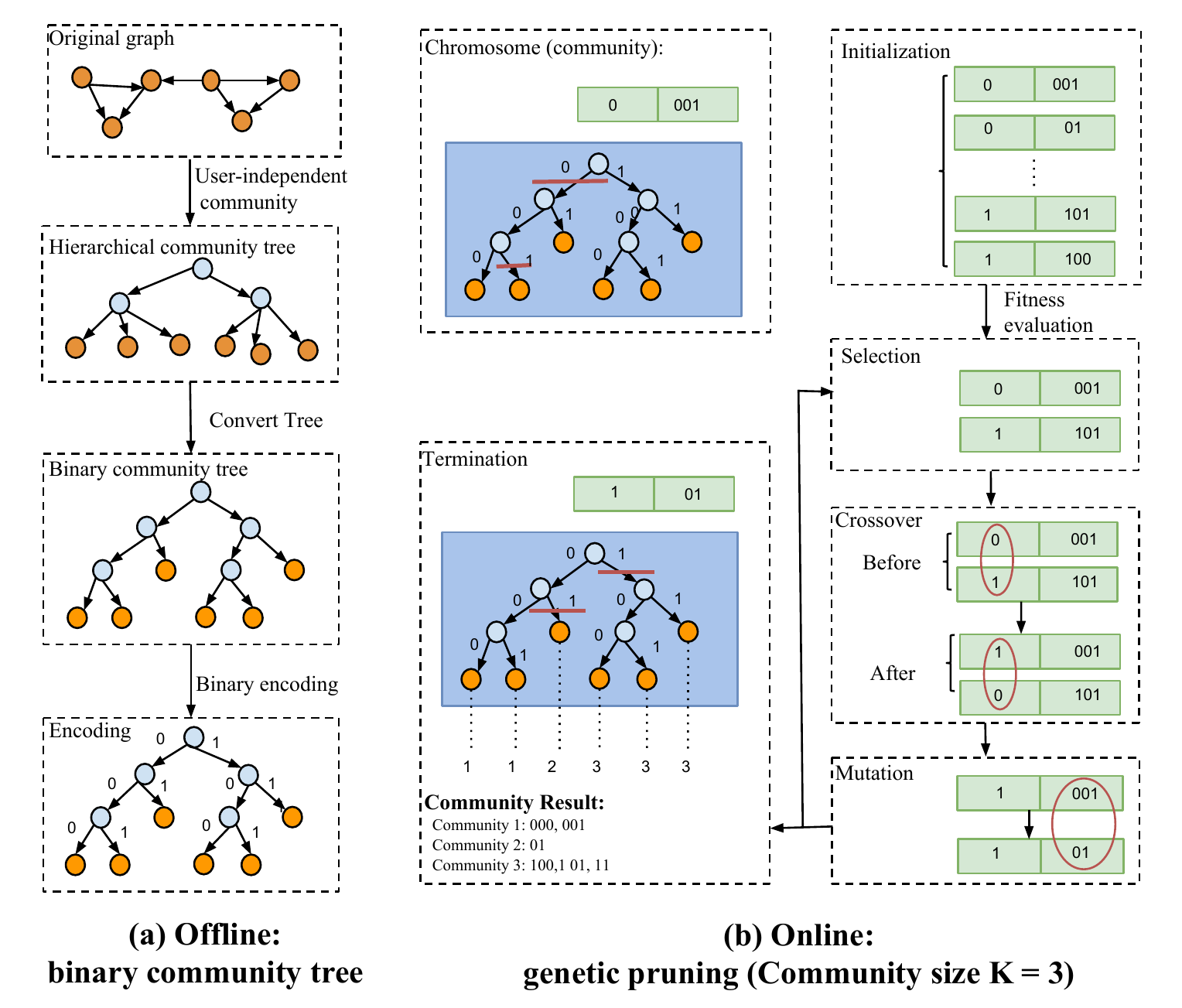}
 \caption{\textbf{The framework of gPCD model}. \textmd{(a) refers to the offline construction step on the original graph and (b) refers to the online genetic pruning step  to generate personalized communities.}} 
 \label{fig:pipeline}
\end{figure}

Our proposed gPCD model contains an offline and an online step. Figure \ref{fig:pipeline} shows the pipeline of the whole framework. The offline step first encodes the user-independent binary community tree (\textit{Section 3.1}), and subsequently learns embedding representations for both user need and nodes on the binary community tree (\textit{Section 3.2}). The online step introduces the genetic personalized community detection approach (\textit{Section 3.3}). To accelerate the running speed, a distributed version of gPCD model is also deployed on HDFS (\textit{Section 3.4}).  To disambiguate the notations mentioned in this section to better explain our gPCD model, some commonly used notations can be found in Table \ref{tab:notation}. 
 
\subsection{Offline Community Tree Index} 
 
One challenge of solving personalized community detection problem is the computational cost due to the complexity of personalization and graph structure. In order to reduce the online workload, most of computation cost is put into the one-time offline step whose time cost can be excluded from the online personalized community detection step. Thus, we first convert the graph into a binary community tree offline to retain user-independent community information. 

\begin{table}[h]
\small
\centering

\begin{tabular}{p{1.5cm}p{6cm}} 
  \toprule 
   \textbf{Notations} & \textbf{Descriptions} \\ \midrule 
   $G(V,E)$ & Original graph $G$ with vertex set $V$ and edge set $E$ \\ \midrule 
   \textit{$T_{c}(N^{c},L^{c})$} & The hierarchical community tree generated from graph $G(V,E)$ with node set $N^{c}$ and link set $L^{c}$. Each node  $N^{c}_{k} \in N^{c}$ denotes a group of vertices belonging to $V$. \\ \midrule 
    \textit{$T_{b}(N^{b},L^{b})$} & The binary community tree reconstructed from \textit{$T_{c}(N^{c},L^{c})$}. Each node $N^{b}_{k} \in N^{b}$ denotes a group of vertices belonging to $V$. \\ \midrule 
   $B$ & The binary codebook for \textit{$T_{b}(N^{b},L^{b})$}. Particularly, $B_{k} \in B$ denotes the binary code of both $N_{k}^{b} \in N^{b}$ and $L_{k}^{b}\in L^{b}$ where $L_{k}^{b}$ is the link points to node $N_{k}^{b}$.
  \\ \bottomrule
\end{tabular}
\caption{Commonly used notations in gPCD model}
\label{tab:notation}
\vspace{-1em}
\end{table} 

We employed the Infomap algorithm \cite{rosvall2011multilevel} to generate the user-independent communities solely based on the graph \textit{G(V, E)}. Infomap algorithm simulates a random walker wandering on the graph and indexes the  description length of his random walk path via multilevel codebooks. By minimizing the description length based on the map equation below, community structures are formed for the graph.
\begin{equation} \small
\textit{$L(M)$} = q_{\curvearrowright}H(\mathcal{Q})+\sum_{i=1}^{m} \textit{$p_{\circlearrowright}^{i}$}H(\textit{$\mathcal{P}^{i}$}) 
\end{equation}
where \textit{$L(M)$} is the description length for a random walker in the current community $M$. $q_{\curvearrowright}$ and $p_{\circlearrowright}^{i}$ are the jumping rates between communities and within the $i_{th}$ community. $H(\mathcal{Q})$ is the frequency-weighted average length of codewords in the global index codebook and $H(\mathcal{P}^{i})$ is frequency-weighted average length of codewords in the $i_{th}$ community codebook. Followed by this equation to partition communities into sub-communities , a hierarchical community tree \textit{$T_{c}(N^{c},L^{c})$} is constructed from the original graph $G(V,E)$. 

In \textit{$T_{c}(N^{c},L^{c})$}, each parent node can have multiple child nodes which can be regarded as a community partition on the parent node. For instance, a node $N_{k}^{c} \in N^{c}$ from $T_{c}(N^{c},L^{c})$ represents a community of vertices.  Its $m$ child nodes $\{N_{k_{1}}^{c},N_{k_{2}}^{c},...,N_{k_{m}}^{c}\}$ represent $m$ sub-communities of vertices from $G(V,E)$ where we have $	\bigcap_{i=1}^{m} N_{k_{i}}^{c} = \varnothing$ and  $	\bigcup_{i=1}^{m} N_{k_{i}}^{c} =  N_{k}^{c}$.  

In order to achieve an efficient personalized community detection in the following online step, we convert the hierarchical community tree \textit{$T_{c}(N^{c},L^{c})$} to a binary community tree \textit{$T_{b}(N^{b},L^{b})$} for index. Specifically, for $m$ child nodes of a parent node $N_{k}^{c}$, a bottom-up approach is proposed to merge a selected pair of sibling nodes as a new node in an iterative manner. The approach runs until all $m$ child nodes merged together to form the parent node $N_{k}^{c}$. To avoid an unbalanced tree where small communities are always left to merge with huge communities in the end, we first select the node with the smallest community size among all sibling nodes in each merging step. It is merged with its sibling node with the largest normalized linked weight (Please refer to Figure \ref{fig:pipeline}(a)). The normalized linked weight function $w(\cdot)$ between  two nodes $N_{i}^{c}$ and $N_{j}^{c}$ is defined as:
\begin{equation} 
\textit{$w(N_{i}^{c},N_{j}^{c})$} =\frac{\textit{$N_{i}^{c}\odot N_{j}^{c}$}-\frac{\mathcal{D}(N_{i}^{c})\cdot \mathcal{D}(N_{j}^{c})}{2\lVert E \rVert} }{\textit{$\lVert N_{i}^{c}\rVert\lVert N_{j}^{c}\rVert$}} 
\end{equation} 
where \textit{$N_{i}^{c}\odot N_{j}^{c}$} denotes the number of edges linked between vertices in node \textit{$N_{i}^{c}$} and \textit{$N_{j}^{c}$}, which can be interpreted as the linkage strength between them; \textit{$\lVert  N_{i}^{c}\rVert$} is the number of vertices inside node \textit{$N_{i}^{c}$}; \textit{$\mathcal{D}(N_{i}^{c})$} is the out-degree of node \textit{$N_{i}^{c}$} (the total number of edges linked to other nodes) and \textit{$\lVert E\rVert$} is the total number of edges in the original graph $G(V,E)$ . $\frac{\mathcal{D}(N_{i}^{c})\cdot \mathcal{D}(N_{j}^{c})}{2\lVert E \rVert} $ denotes the random linkage strength between node \textit{$N_{i}^{c}$} and \textit{$N_{j}^{c}$}. The \textit{$w(\cdot)$} function calculates how much that two nodes are better connected beyond random connection and is normalized by node size. Given the node $N_{i}^{c}$ with the smallest community size and all its sibling node set $S$, The merging step can be formulated as:
\begin{equation}  
\begin{aligned} 
& N_{j}^{c} \Leftarrow \argmax_{N_{j}^{c} \in S}w(N_{i}^{c},N_{j}^{c})\\
& N_{*}^{c} = N_{i}^{c}\bigcup N_{j}^{c}
\end{aligned}
\end{equation}
 The bottom-up process will stop until all child nodes are merged together to form the parent node. In the end, the hierarchical community tree \textit{$T_{c}(N^{c},L^{c})$} is fully converted to a binary community tree \textit{$T_{b}(N^{b},L^{b})$} with user-independent community information.  The node size $\lVert  N^{b}\rVert$ as well as the link size $\lVert  L^{b}\rVert$  in \textit{$T_{b}(N^{b},L^{b})$} is at most $2\lVert V \rVert$ which is smaller than the size of original graph $G(V,E)$. If we consider to form the binary community tree with only $k$ levels, the size of \textit{$T_{b}(N^{b},L^{b})$} can be even smaller. 
 
 For running time analysis, calculating normalized linked weight takes constant time. In each merging step, node pair selection takes linear time. Therefore, in the worst case, the time complexity of binary community tree construction is $O(\rVert V\lVert^2)$ where the depth of the hierarchical community tree $T_{c}(N^{c},L^{c})$ is 1 and each vertex in $G(V,E)$ forms a single-vertex community.
  
To encode the nodes and links on $T_{b}(N^{b},L^{b})$ as binary code, the root node is encoded as `null' first. For a parent node \textit{$N_{k}^{b}$} with its left child node \textit{$N^{b}_{k_{l}}$} and right child node \textit{$N^{b}_{k_{r}}$}, the binary code of a child node and the related link defined in the Notation Table \ref{tab:notation} is calculated as: 

\begin{eqnarray}\text{$B_{k_{i}}$}=
\begin{cases}
\text{$B_{k}$}+``0", & i = ``l"\cr 
\text{$B_{k}$}+``1", & i = ``r"
\end{cases}
\end{eqnarray} 
 
For instance, if the node \textit{$N_{k}^{b}$} is with binary code ``$00$," its left child node's binary code is ``$000$" while the right child node's binary code is ``$001$." The link \textit{$L_{k}^{b}$} that points to \textit{$N_{k}^{b}$} also has the binary code ``$00$".


\subsection{Community and User Need Representation}

Node2vec \cite{grover2016node2vec} helps to learn fixed-length embeddings for both user need and communities. It simulates random walks on the graph $G(V,E)$ and learns the vertex embedding by optimizing the sequential relationships from random walk paths. In the end, each vertex $V_{k}$ in graph $G(V,E)$ has a vector representation as $\vec{V_k}$. Each node $N_{k}^{b}$ on the binary community tree \textit{$T_{b}(N^{b},L^{b})$} refers to a vertex community $C_{k}$ in the graph $G(V,E)$. Its representation $\vec{C_{k}}$ is calculated as the averaged embedding of all vertices inside the community. In the end, the binary community tree \textit{$T_{b}(N^{b},L^{b})$} represents the hierarchical community partition of Graph $G(V,E)$. Each node \textit{$N_{k}^{b}$} on the tree is indexed with three attributes: a group of vertices from graph $G(V,E)$, a binary code \textit{$B_{k}$}, and an embedding representation \textit{$\vec{C}_{k}$}.

On the other hand, User need (query) $I$ can also be represented by a combination of $t$ different vertices $\{V_1, V_2... V_t\}$ in the graph $G(V,E)$. In this study, two different scenarios for user need representation are offered:

\textbf{Vertex-based Query}. User need can be directly represented by the vertices based on the generation probability $P(V_k|I)$ between them. Hence the user need representation $\vec{I}$ is calculated as:

\begin{equation}
\vec{I} = \sum_{k=1}^{t} P(V_k|I) \cdot \vec{V_k} 
\end{equation} 
For instance, in a music sharing network, each vertex $V_k$ denotes a music and a user listing history can be used to reflect the user need $I$. $P(V_k|I)$ therefore can be regarded as the probability that a music being listened by the user. 

\textbf{Text-based Query}. Under this scenario, user need $I$ is represented as a text query, and each vertex $V_{k}$ in the graph $G(V,E)$ also contains textual content. From language model viewpoint, each vertex importance weight is the query likelihood $P(I|V_k)$, and the user need can is the weighted average of vertex embedding: 

\begin{equation}
\vec{I} = \frac{\sum_{k=1}^{t} P(I|V_k) \cdot \vec{V_k}}{\sum_{k=1}^{t} P(I|V_k)}
\end{equation}

In either case, user need is conceptualized as an embedding with the same dimension as the node embeddings on the binary community tree. It enables very efficient online personalized community detection in later steps. And running Node2vec takes most of the time in this step.

\subsection{Online Genetic Pruning}

The whole process, as the Figure \ref{fig:pipeline} shows, is to generate communities by pruning the constructed binary community tree. After each cut on a link, the original tree  will be separated into two sub-trees. After a specific number of cuts to the links on the tree, a fixed number of communities with different resolutions are detected.  By applying genetic selection, crossover, and mutation steps, our model converges to the optimized solution efficiently with a clear-defined fitness function. The details are shown in the following paragraphs.

\subsubsection{Genetic Representation} \ 

A chromosome is formed by a set of genes \{$g_{1},g_{2},...,g_{K-1}$\}, and each gene $g_{i}$ holds a cut link $L_{i}^{b}$ in the binary community tree \text{$T_{b}(N^{b},L^{b})$}. Since communities can be created by cutting links on the offline tree, a chromosome can be represented as a generated community partition of the original graph $G(V,E)$ in this way. 
To constrain a chromosome so that it can be decoded to a fixed number of communities, four \textbf{Cutting Rules} are necessarily to be applied: 
\begin{itemize} 
 \item \textbf{ Rule 1}: If a link $L_{i}^{b}$ is picked to cut on the binary community tree \text{$T_{b}(N^{b},L^{b})$}, its pointing node $N_{i}^{b}$ will be retrieved and all the vertices within it form a community. 
 \item \textbf{ Rule 2}: If a link {$L_{i}^{b}$} and its ancestor link {$L_{j}^{b}$} are stored in the same chromosome, all vertices in {$L_{i}^{b}$}'s related node {$N_{i}^{b}$} are a subset of vertices in {$L_{j}^{b}$}'s related node {$N_{j}^{b}$}. In this case, the two cut links generate two communities where community $C_{i}$ is all vertices in $ N_{i}^{b}$ and community $C_{j}$ is the remaining vertices in $ N_{j}^{b}$ but not in $ N_{i}^{b}$. It can be formulated as   $C_{i} = \bigcup_{k}\{V_{k}|(V_{k}\in N_{i}^{b})\}$ and community $C_{j} = \bigcup_{k}\{V_{k}|(V_{k}\in N_{j}^{b}) \cap (V_{k}\notin N_{i}^{b})\}$.  
 \item \textbf{ Rule 3}: Sibling links can't be stored in the same chromosome, and it is not allowed to store duplicated links in a chromosome.
  \item \textbf{ Rule 4}: The depth's upper bound is set to be $d$, which means all eligible cut links should be located in the first $d$ depth on the binary community tree. It avoids to generate super tiny communities and hugely reduces the genetic searching scope on cut links.
\end{itemize} 
 
By applying the cutting rules to the online pruning process, we ensure a $K$ community partition can be retrieved from a chromosome with $K-1$ cut links. 

\subsubsection{Initialization} \ 

Initially, the model generates a given number $P$  chromosomes as the seed ``chromosome population". And each iteration in the genetic approach breeds a new ``generation'' of chromosome population. In order to ensure a chromosome is an encoder of a $K$ community partition, $K-1$ links will be randomly picked (on the binary community tree) following the cutting rules and stored in the related genes of a chromosome. 

\subsubsection{Fitness Function} \ 

As each chromosome can be decoded as a community partition, it is important to measure the quality of each generated chromosome (how well the generated communities can satisfy user need). The measurement is hosted in a fitness function. 



In our model, the fitness function simulates the user searching behavior on the graph given the community partition. For instance, a user can be more likely to pick the most relevant communities while avoiding the redundant information already selected. With the help of the offline step, the relevance score of node $N_{i}^{b}$ (community $C_{i}$) towards user need $I$ can be calculated with the cosine similarity $cos(\vec{I},\vec{C_{i}})$, and the information redundancy can be $\sum_{C_{j} \in S_c}cos(\vec{C_{j}},\vec{C_{i}})$ where $S_{c}$ is the set of communities that the user have already picked from the communities decoded from the target chromosome. Following this, we use a greedy selection approach to iteratively rank and pick communities given a chromosome (community partition) until all communities are picked: 




\begin{equation}  \argmax_{C_{i}}\lambda \cdot cos(\vec{I},\vec{C_{i}})-(1-\lambda)\cdot \frac{\sum_{C_{j} \in S_c}^{ }cos( \vec{C_j},\vec{C_{i}})}{\lVert S_c\rVert}
\end{equation} 

where $C_i$ is the candidate community to be picked and $\lVert S_c\rVert$ is the number of communities already been picked. $\lambda$ is a parameter controls  whether user prefers to obtain new useful information or to avoid redundant information. 

For chromosome quality evaluation, a query-generated vertex ranking list $l_{q}$ is first created by retrieving top $n$ vertices relevant to the query (user need) with the largest cosine similarities on embeddings of graph $G(V,E)$. We store the top $n$ vertex ranking label $R(l_{q})=\{1,2,...,n\}$ as the pseudo ground truth. On the other hand, given the $k_{th}$ chromosome $ch_k$ in the current chromosome generation, we can also retrieve the community-generated ranking of each vertex $V_{k} \in l_{q}$ from the sequentially selected communities decoded by the chromosome. We assign the ranking label on each vertex $V_{k}$ based on the following formula:
\begin{equation}
	\sum_{V_{j} \in l_{q}}\Phi(\delta(V_{j}) < \delta(V_{k})) + 1
\end{equation} 


$V_{j}$ refers to all vertices in $l_{q}$. $\delta(V_{j})$ shows the ranking (selection sequence) of the community which $V_{j}$ belongs to. $\Phi$ is a binary operator to determine whether $V_{j}$ satisfy the condition $\delta(V_{j}) < \delta(V_{q})$. This formula helps to construct the community-generated ranking label $R(l_c)$. For instance, when $n=3$, we have a query-generated ranking list $l_q = \{V_1,V_2,V_3\}$ and its related ranking label $R(l_q) = \{1,2,3\}$. Given a chromosome where the community of $V_1$ and $V_2$ is the same and selected before $V_3$, we can generate the related community ranking label $R(l_c) = \{1,1,3\}$ with the same vertex sequence of $l_q$.

Then, we define the fitness function $f(\cdot)$ to evaluate the chromosome $ch_{k}$. As we have the query-generated ranking label $R(l_{q})$ (ground truth) and community-generated ranking label $R(l_{c})$ from $ch_{k}$, we calculate their Kendall's $\tau$  correlation coefficient  as the fitness score $f(ch_k)$ of chromosome $ch_k$ where higher score means the chromosome $ch_{k}$ can  generate better personalized communities to meet with user need. 

\begin{equation}
\textit{$f(ch_{k})$} = 1-\frac{\sum_{i=1}^{n} R(l_{ci})\cdot R(l_{qi})}{\sum_{i=1}^{n} R(l_{ci})^2\cdot \sum_{i=1}^{n}R(l_{qi})^2}
\end{equation} 
where $R(l_{ci})$ is the $i_{th}$ vertex ranking in community-generated ranking label $R(l_{c})$ and $R(l_{qi})$ is the $i_{th}$ vertex ranking in query-generated ranking label $R(l_{q})$. Kendall's $\tau$ is a widely used metric to evaluate the correlation between two lists where higher score means stronger correlation. Thus, higher fitness score reflects that the generated community ranking ($R(l_{c})$) can better meet with user need ($R(l_{q})$). 

Moreover, it is clear that the fitness function aims to separate all top $n$ vertices in different communities to get the optimal case. It matches our research goal to generate high resolution communities on vertices which are more relevant to user need. As the number of community is a given number $K$, it also leads to a coarser manner partition on the remaining less relevant vertices. On the other hand, the binary community tree $T_{b}(N^{b},L^{b})$ and the Cutting Rule 4 naturally preserve the community structure and unite the most relevant vertices in the same community. Hence the whole genetic approach is a gambling process. The final chromosome result is the equilibrium case to detect communities both contain graph topological structure and meet with user need.


\subsubsection{Selection}\ 

We select the superior chromosomes from current chromosome population based on their fitness scores. The probability that the $k_{th}$ chromosome $ch_k$ is picked can be calculated via the Softmax normalization function $p(ch_{i}) = \frac{exp(f(ch_{k}))}{\sum_{i=1}^{P}exp(f(ch_{i}))}$. Then, the Fitness Proportionate Selection method \cite{fogel1997evolutionary} is applied to randomly select $P$ chromosomes into chromosome pairs based on probability distribution. In order to enhance optimization efficiency, we also use elitism selection to ensure the best chromosome in the current generation will always be selected to the next generation. 

\subsubsection{Crossover}\ 

To reach global optimum community partition efficiently, given a pair of chromosomes, the crossover operation can randomly exchange part of the genes in both chromosomes to produce a new pair of chromosomes with a certain crossover rate.

In order to make sure that the newly generated chromosomes meet the cutting rules, an \textbf{Exchange Rule} is defined to restrict gene exchange: If gene \textit{$g$} contains link {$L^{b}_{g}$}, \textit{$g$} can't do crossover process with genes that contain either link {$L^{b}_{g}$} or its sibling link {$L^{b'}_{g}$}. This rule can help avoid having duplicated links or sibling links stored together in the newly generated chromosome (To satisfy Cutting rule 3). 

After \textit{$m$} random numbers are selected from \{$1,2,...,K-1$\} as exchanged gene position indexes, genes located in the chosen positions of two chromosomes will exchange the stored link restricted by the Exchange Rule.

\subsubsection{Mutation} \ 

Mutation operation is applied to avoid local optimization. If a chromosome is chosen to mutate, a gene within the chromosome will be randomly picked, and its stored link will be changed to another link restricted by the Exchange Rules. An example is illustrated in Figure \ref{fig:pipeline}(b) where the link stored in the second gene is changed from ``001'' to ``01''.





\subsubsection{Termination}\ 

After $T$ iterations, the whole process stops and the current best chromosome is retrieved as the final result. Choosing the number of $T$ is dependent on the task. In order to decode the final chromosome to the related community partition, all genes in the chromosome are sorted in an ascending order based on the binary code of their stored cut link. Vertices whose binary codes start with the same cut link's binary code will be assigned to the same community label. And its later assigned community label can overwrite the previous assigned community label. For instance, if there are a vertex with binary code ``0011'' and two cut links with binary code ``00'' and ``001'', the vertex will be assigned to a community label ``00'' first, and its community label is overwritten by ``001'' afterwards. The Termination step in Figure \ref{fig:pipeline}(b) also illustrates a vivid example. In this way, the binary code of the binary community tree can help to decode the final chromosome into communities in an efficient way.


\subsection{Distributed gPCD}
\begin{figure}  
 \center
 \includegraphics[width=1\columnwidth]{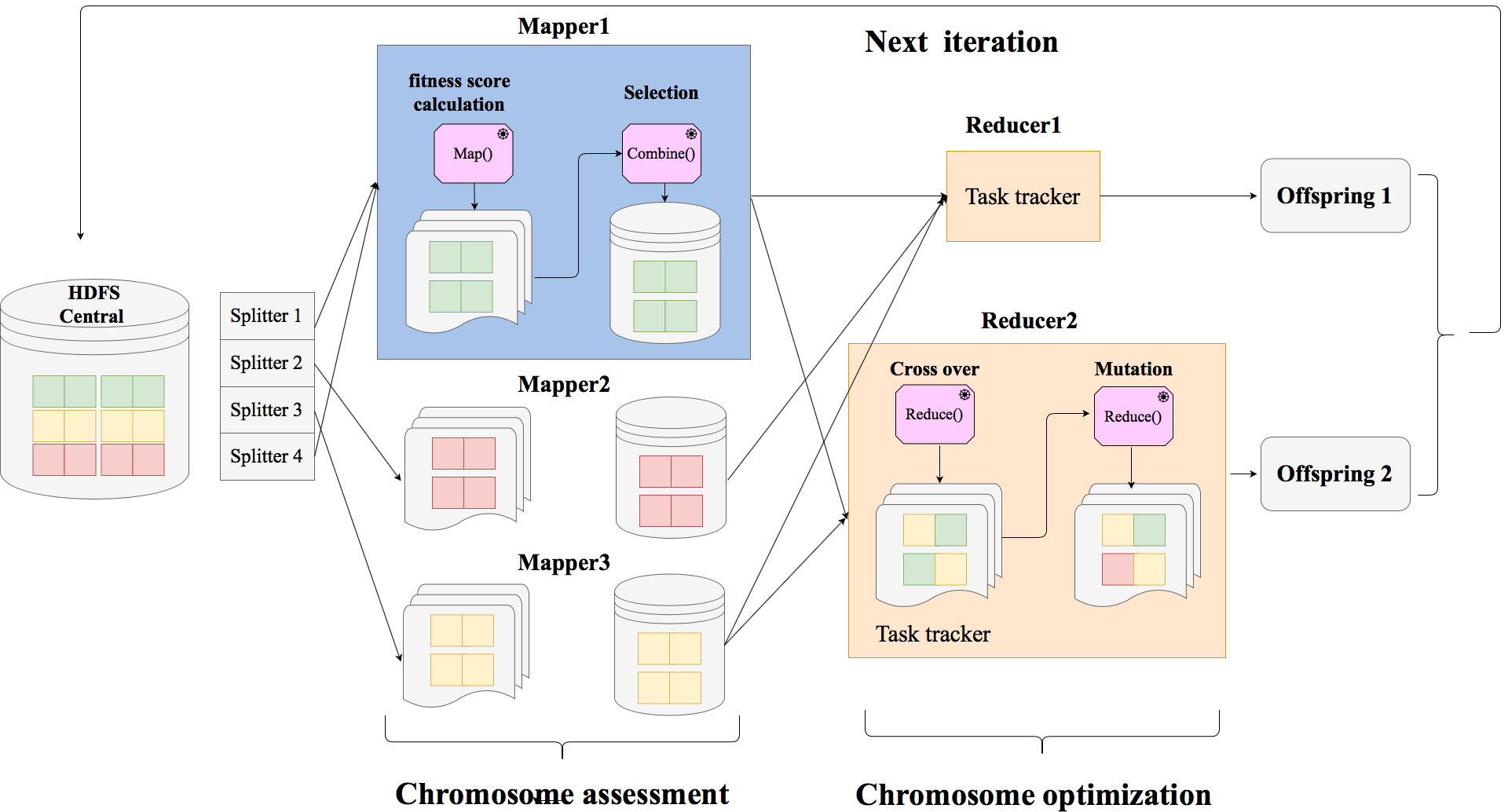}
 \caption{Online parallel computing process on Hadoop Distributed File System (HDFS)}
 \label{fig:distributed}
\end{figure}

To enhance the online step efficiency, a MapReduce framework \cite{ferrucci2013framework} is utilized to enable the distributed genetic evolution. Figure \ref{fig:distributed} depicts the personalized community detection under a MapReduce framework. The chromosome collection is either originally initialized from binary community tree or obtained from the last generation. It contains the whole chromosome population in the central depository. In its first ``Splitter'' process, all chromosomes are split into $M$ groups based on their hash values and sent out to related $M$  Mappers to calculate the ``Fitness'' scores. In the same Mapper, after all chromosomes are assigned fitness scores, based on their scores, a Combiner groups all chromosomes together and random select equal number of chromosomes with duplicated as the ``Selection'' step. All the selected chromosomes are sent to $R$ reducers (we set $R = M$ arbitrarily in order to better represent time complexity) to form pairs for the ``Crossover'' and  ``Mutation'' step, calculate new chromosome offsprings for the next generation and store them back to the central repository. 


The complexity of the proposed algorithm is $O(2^dKP)$ without parallel computing and $O(\frac{2^dKP}{M})$ with  parallel computing, where $d$ denotes the upper bound where the cut links are restricted in the top $d$ depth of the binary community tree $T_{b}(N^{b},L^{b})$; $K$ denotes the community number; $P$ denotes the initialized population size of the genetic algorithm and $M$ denotes the number of Mappers/Reducers in parallel environment. As all the parameters are considerably small (compared with the node/edge size in the original graph), the whole process runs very fast to retrieve the final community partition.

%% file: content/experimentRevised.tex
\section{Experiments}
 \begin{figure*}
	\subfloat[Depth $d$ in Citation model]{\includegraphics[width = 1.4in]{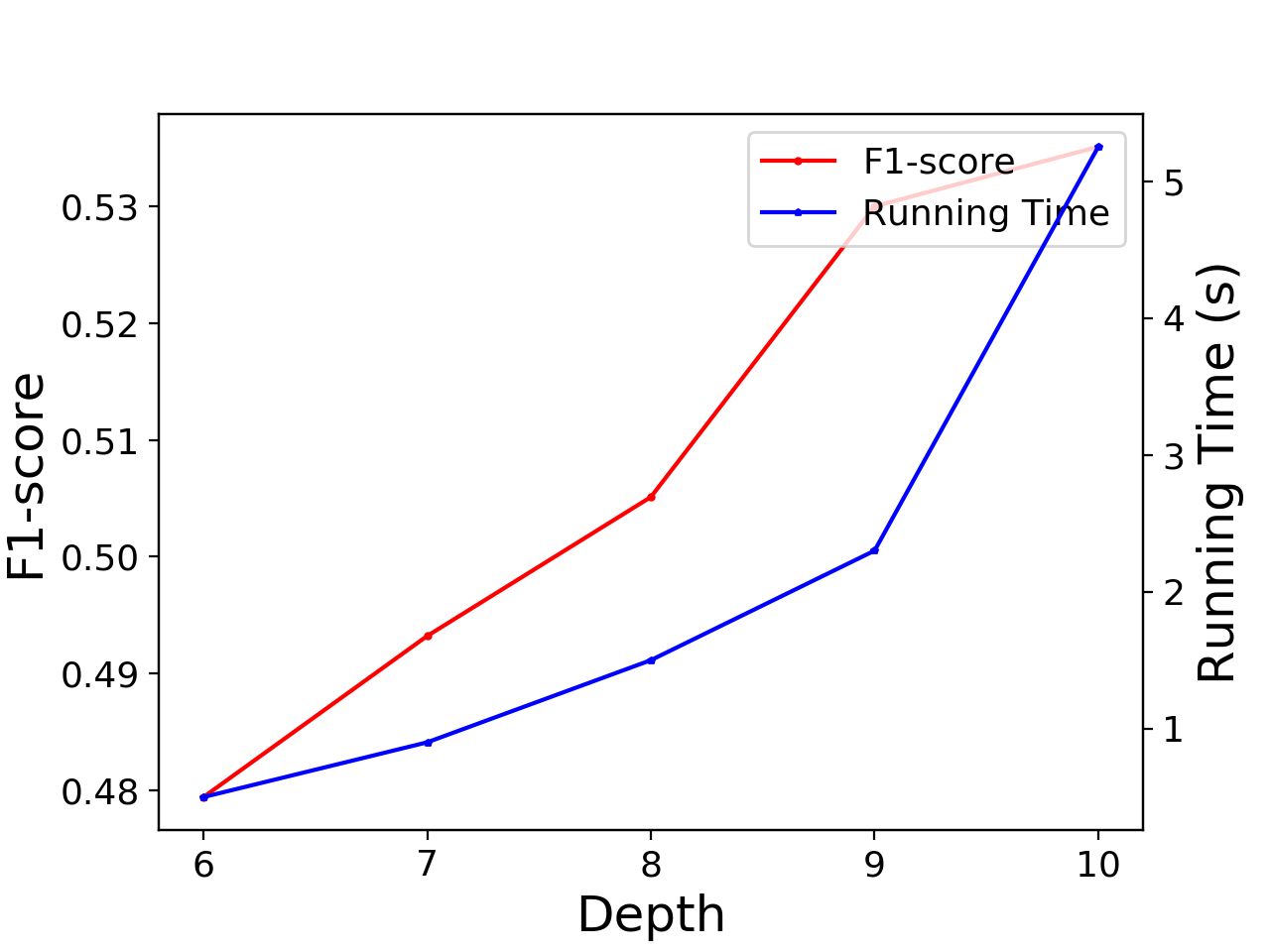}}
	\subfloat[Depth $d$  in Keyword model]{\includegraphics[width = 1.4in]{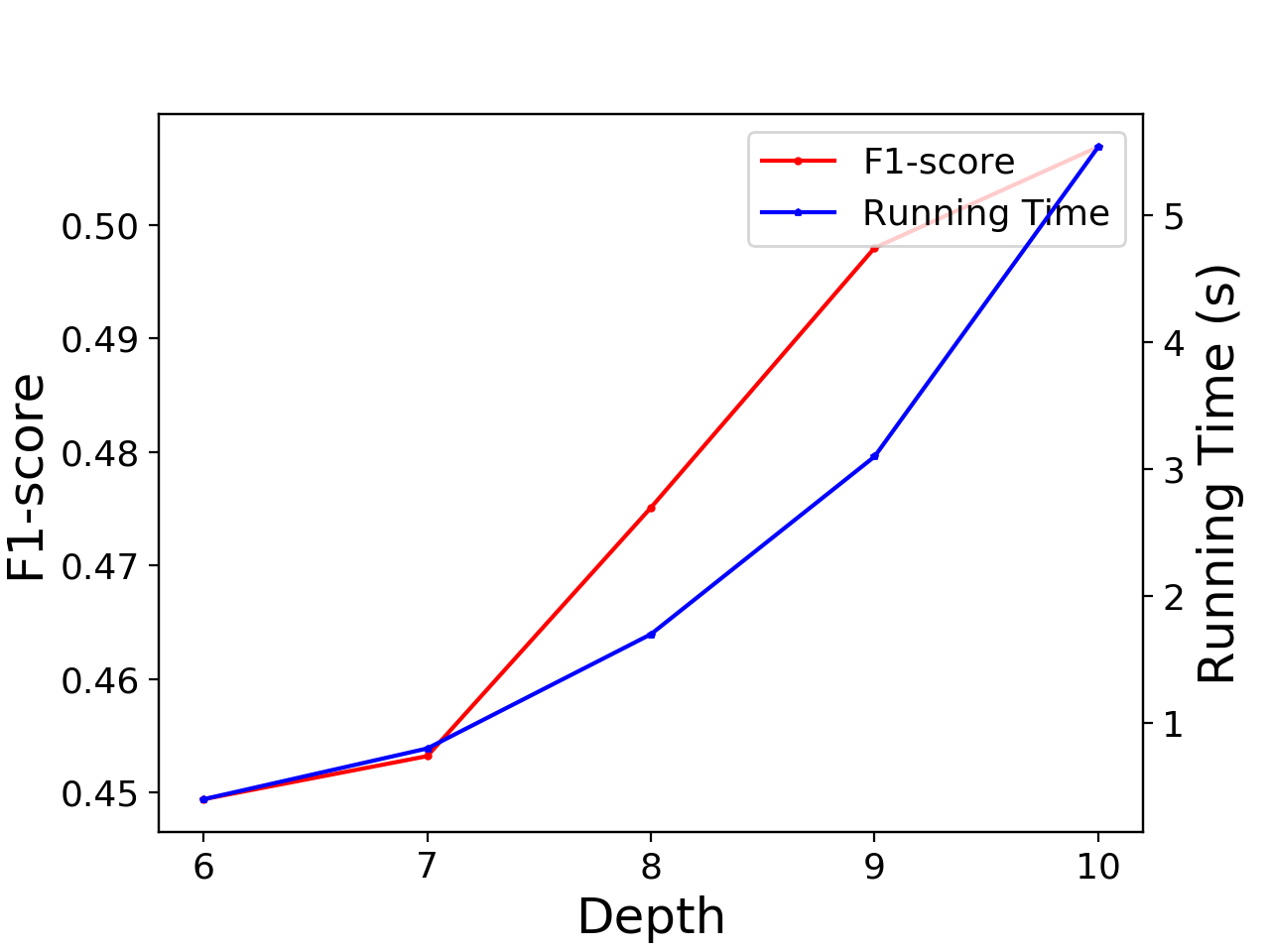}} 
	\subfloat[Depth $d$  in Listening model]{\includegraphics[width = 1.4in]{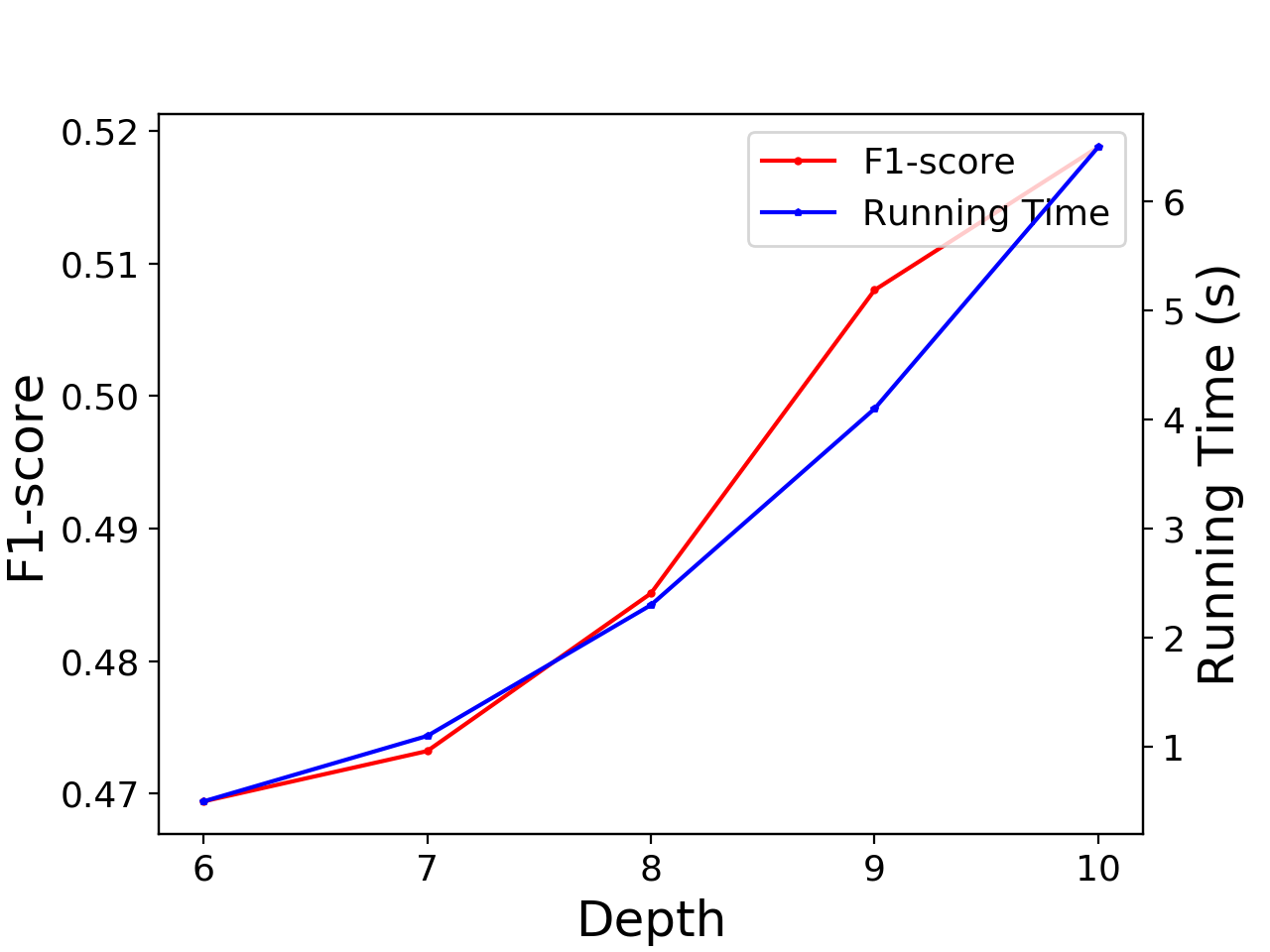}}
	\subfloat[Iteration $T$ in all models]{\includegraphics[width = 1.4in]{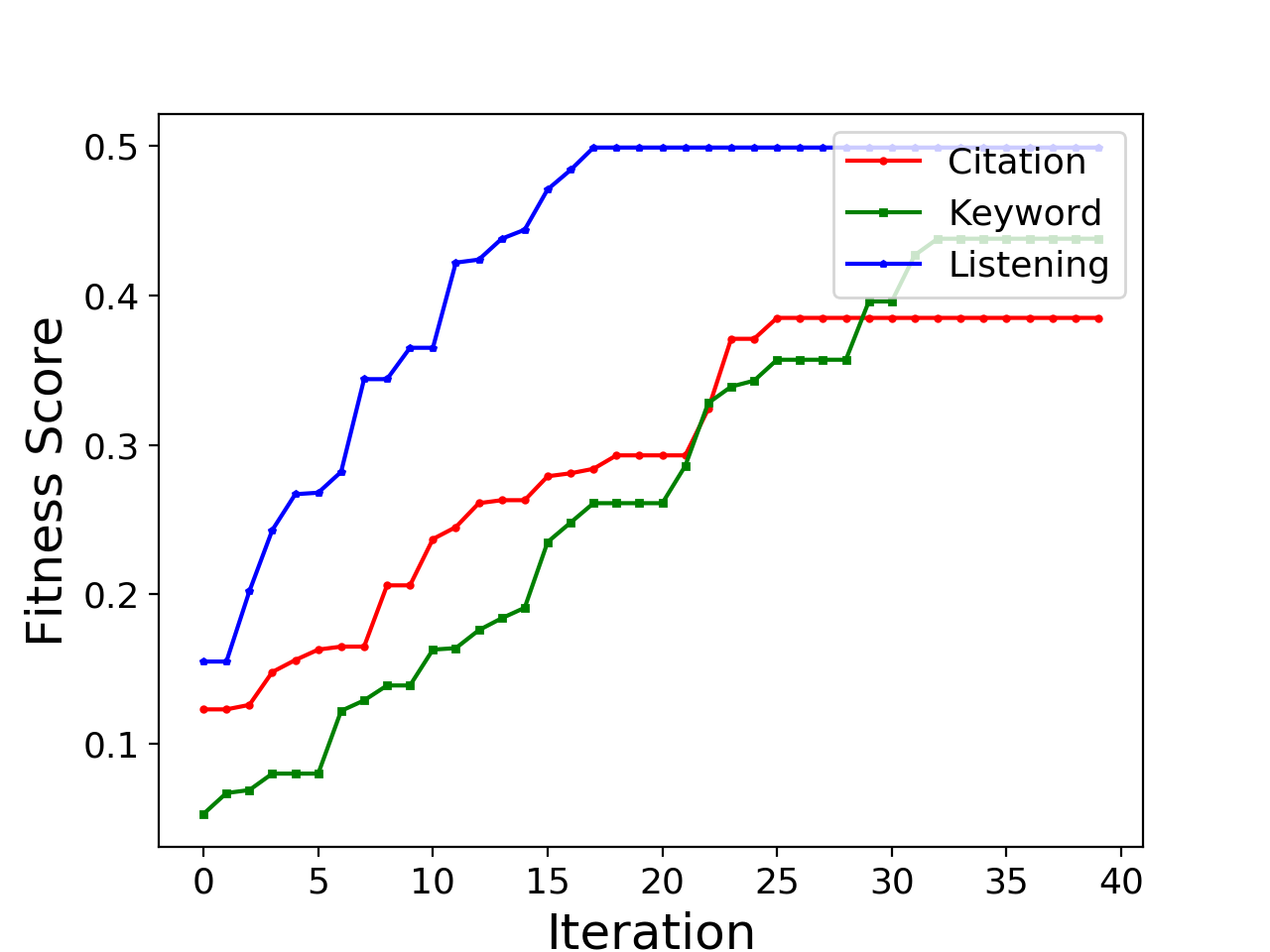}} 
	\subfloat[User searching preference $\lambda$ in all models]{\includegraphics[width = 1.4in]{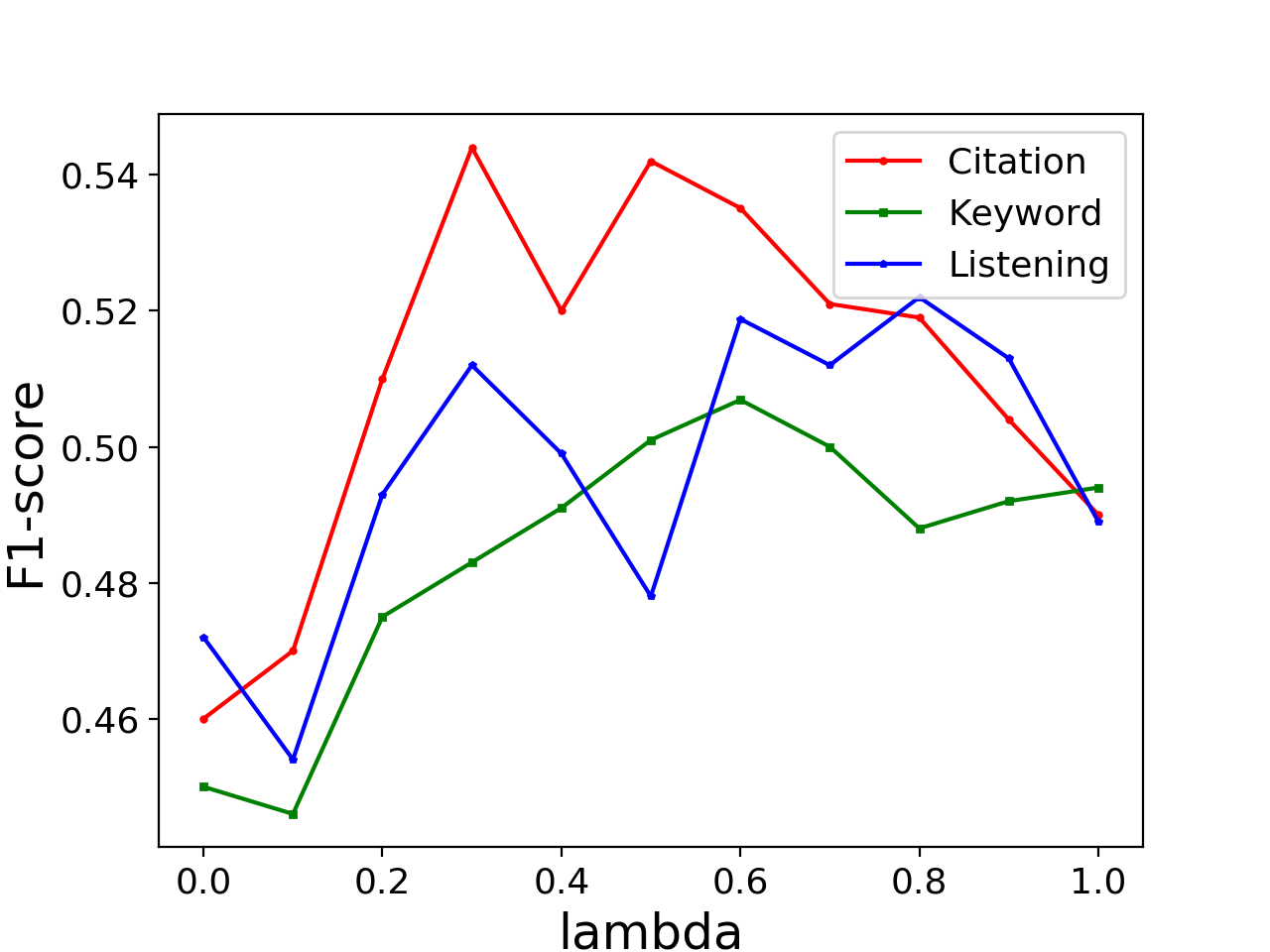}} 
	\vspace{-1em}
	\caption{Parameter effects on model performance}
	\label{fig:tuning}
	 \vspace{-1em}
\end{figure*}

\subsection{Datasets} 
\subsubsection{Datasets Description} \ 

Table \ref{tab:dataset} shows the statistics of the two datasets.  The scholarly graph is unweighted and directed, while the music graph is a weighted and undirected.


\begin{table}[h]
	\centering
	
	\begin{tabular}{ccccc} 
		\toprule
		\multirow{2}{*}{\textbf{Dataset}}  &\multicolumn{2}{c}{\textbf{Node description }} & \multicolumn{2}{c}{\textbf{Edge description}}\\ \cmidrule(lr){2-5}
		& Type & Size & Type & Size \\ \midrule 
		Scholarly & paper & 166,170 & citation &750,181\\ \midrule 
		Music & song & 145,203 & co-listening &1,172,525
		\\ \bottomrule
	\end{tabular}
	\caption{Dataset Description}
	\label{tab:dataset}
\end{table} 
\textbf{Scholarly Graph:} It contains academic publications with metadata extracted from ACM Digital Library. From the dataset, we build the experimental graph via paper citation relationship. Each vertex in the graph represents a paper, and if a paper cites another paper, there will be an edge linking the two. Our model aims to detect personalized communities on the scholarly graph for authors. For each author in the dataset, we represent his/her need in two ways: their previous publications (text-based query) and cited paper history (vertex-based query). 

\textbf{Music Graph:} It contains user listening histories and user-generated playlists from an online music streaming service, Xiami. We create a music graph with songs as the vertices and co-listening relationship as the edges. If two songs appear in the same user's listening history, there will be an edge linking them two. For users in the music dataset, we represent their music tastes (user need) from the songs in their listening history. 

 
\subsubsection{Ground Truth Construction} \ 

The ground truth for the two datasets are generated based on each user's publishing/ citing/ listening history:

For the scholarly dataset, the references of 112 random sampled papers are manually annotated from their literature reviews where authors summarize previous works. Different paragraphs (or sub-sections) in the literature review typically focus on separate but coherent topics while the same paragraph talks about the same topic \cite{zhang1916synopsizing}. Based on this assumption, the papers cited in the same paragraph/sub-section naturally form a community with high resolution. To ensure each paper's cited papers form enough communities and each community contains enough papers, only papers with no fewer than three topics and all of whose communities have at least five papers are kept. After applying all these filters, 101 papers are left for evaluation. 

For the music dataset, each user has several self-generated playlists. The songs in each playlist should contain a coherent theme. To avoid the playlists sharing mutually exclusive themes with other playlists, a Jaccard similarity check is applied on any pair of playlists created by the same user. If a user has at least two highly correlated playlists (Jaccard coefficient between them is above 0.5), we remove one of the playlists. Each playlist forms a separate community. Furthermore, to ensure the number of playlist and playlist size are both large enough, only users with at least three playlists and each playlist contains at least five songs are kept. In the end, there are 117 users who meet the above criteria. 

In this paper, as all communities constructed in the ground truth are relevant communities with high resolution for users, our task is generating personalized communities to reconstruct the ground truth on two different datasets with vertex- and text-based user need. We share our code in Github\footnote{https://github.com/RoyZhengGao/gPCD}.
\subsection{Settings}

\subsubsection{Metrics \& Parameter Settings} \ 

F1-score (F1), Rand index (Rand), Jaccard index (Jaccard) and running time are reported as the evaluation metrics in this paper. Based on empirical studies, population size $P$ is 100. Crossover rate is 95\%. Mutation rate is 1\%. The maximum depth of binary community tree $d$ is 10. The number of iteration $T$ is 30. User searching preference $\lambda$ is 0.6. Community size $K$ is 50. The number of Mappers/ Reducers for parallelization $M$ is 50. The number of top vertices to construct pseudo ground truth $n$ is 10. Parameters in Infomap and Node2vec are both the default settings in their original papers.


\subsubsection{Baselines} \ 

Considering both efficacy and efficiency, we select eight widely used user-independent community detection models. Ideally, to achieve personalized community detection, user-independent models should run on each user separately by assigning higher weights on user related edges. Thus, their time complexity should be only compared with our online step time complexity as our offline step is independent with user numbers. In this paper, to run baselines within acceptable time, we report their user-independent community results as the average performance. 

\begin{itemize}
	\item \textbf{Spinglass}: Spinglass \cite{eaton2012spin} constructs communities by minimizing the Hamiltonian score on signed graphs.
	\item \textbf{Fast Greedy} (FG): Fast Greedy \cite{clauset2004finding} is a greedy search method to get the maximized modularity for community detection.
	\item \textbf{Louvain}: Louvain \cite{blondel2008fast} is an agglomerative method to construct communities in a bottom-up manner guided by modularity.
	\item \textbf{Walktrap}: Walktrap \cite{pons2005computing} detects communities based on the fact that a random walker tends to be trapped in dense part of a network.
	\item \textbf{Infomap}: Infomap \cite{rosvall2011multilevel} generates communities by simulating a random walker wandering on the graph and indexing the description length of his random walk path via multilevel codebooks.
	\item \textbf{Bigclam}: Bigclam \cite{yang2013overlapping} generates overlapping communities via a non-negative matrix factorization approach.
	\item \textbf{DeepWalk}: DeepWalk \cite{perozzi2014deepwalk} generates node embeddings via random walks and utilizes K-means on node embeddings to detect communitis.
	\item \textbf{Node2vec}: Node2vec \cite{grover2016node2vec} is an extended version of DeepWalk with a refined random walk strategy.
\end{itemize}
\subsection{Results}

 \subsubsection{Evaluation Results}  \

There are two scenarios to construct user need. For the scholarly graph, including a citation (vertex-based query) model and a keyword (text-based query) model. In the citation model, for each user, we first extract all the papers he/she cited before, then use their centroid embedding as user need vector  $\vec{I}$. In the keyword model, for each user (author), we first extract all keywords he/she used in all previous papers to form a text query. Then we retrieve the top 100 relevant papers given the query based on probability language model with Dirichlet smoothing \cite{larson2010introduction}. Finally, we average those retrieved papers' vectors as the user need vector $\vec{I}$.

 \begin{threeparttable}
	\small
	\centering
	
	\renewcommand{\tabcolsep}{3.0pt} 
	\renewcommand\arraystretch{0.0}
	\begin{tabular}{lcccccc} 
		\toprule   
		\multirow{2}{*}{\textbf{Model}} & \multicolumn{3}{c}{\textbf{Scholarly Graph}} 	&\multicolumn{3}{c}{\textbf{Music Graph}}\\
		\cmidrule(lr){2-7}
		& F1 & Rand	& Jaccard	& F1 & Rand	& Jaccard\\ \midrule
		Spinglass&0.4294 & 0.4149 &0.3593 & 0.4282 & 0.5317& 0.2823 \\ \midrule
		FG& 0.4290 & 0.3852& 0.3735 &0.4645&0.4100& 0.3070 \\ \midrule
		Louvain&0.4417 & 0.4546& 0.3627&0.1832 &0.4201 & 0.1174 \\ \midrule
		Walktrap& 0.4304 & 0.3777 & 0.3777&0.3999 & 0.3507& 0.3490 \\ \midrule
		Infomap&0.4436 & 0.4165& 0.3606& 0.2147 & \textbf{0.6074} &0.1344\\ \midrule
		Bigclam& 0.2314  & 0.2572 & 0.1348 & 0.1499  & 0.2078  & 0.1227 \\ \midrule
		DeepWalk&0.3904 & 0.3237 & 0.3234&0.3535& 0.3253& 0.3001 \\ \midrule
		Node2vec & 0.4001 & 0.3472 & 0.3433&0.4122 & 0.4101& 0.3087 \\ \midrule
		gPCD-Citation&\textbf{0.5351}$^{*}$ & \textbf{0.4551}$^{*}$& \textbf{0.4086}$^{*}$ &-	&-	& -\\ \midrule
		gPCD-Keyword&0.5069 & 0.4114& 0.3708 &-	&-	& - \\ \midrule
		gPCD-Listening&-	&-	& - &\textbf{0.5188}$^{*}$ & 0.5865& \textbf{0.3550}$^{*}$  \\ 
		\bottomrule
	\end{tabular}
	\begin{tablenotes}
		\item[] \begin{footnotesize}
			Note: ``*'' means the p-value through a pairwise t-test is smaller than 0.001.
		\end{footnotesize}
	\end{tablenotes}
	\caption{Personalized Community Evaluation on gPCD and Baselines. }
	\label{tab:community}
\end{threeparttable}

 For the music task, text information is not available. The centroid embedding of the songs listened to by a target user are taken as user need.
 
 \textbf{Community Accuracy}: We compare the average performance of our model on all testing users with baselines. Table \ref{tab:community} shows the detailed metrics. When running on the Scholarly graph, both Citation model and Keyword model can achieve around 10\% increase on F1-score compared with all baselines. Citation model also performs the best in Rand Index and Jaccard Index. For Music graph, our Listening model also has a significant improvement on F1-score and Jaccard Index. Although it has similar performance on Rand Index compared with Infomap, we believe our model in fact works much better due to the Infomap's poor performance on the rest two metrics. Moreover, we apply pairwise t-tests \cite{derrick2017impact} for all metrics on all testing users. All metrics' p-values in Citation model and the p-values of F1-score and Jaccard Index in Listening model are all smaller than 0.001, which means the improvements of our model performance are significant compared with baselines.

 \textbf{Running Time Comparison}: Table \ref{tab:complexity} shows both the theoretical time complexity and real running time. To represent baseline algorithms' time complexity, ``$V$" refers to the vertex number and ``$E$" refers to the edge number in the graph $G(V,E)$. For some models (FG, Walktrap, and Infomap.), their specific time complexities are officially mentioned in the original papers. The time complexity of Louvain and Bigclam are roughly estimated in the original papers as well but those papers don't mention specific numbers. For Spinglass, DeepWalk and Node2vec, we can't find the exact time complexity in existing studies. Hence in this paper, we arbitrarily assign labels based on the their real running speed. Considering the running time, all the baseline algorithms run relatively fast except for the Spinglass algorithm. However, compared with all other models, the distributed gPCD always performs the fastest. Its real running time of is less than one-tenth of the fastest baseline's running time. 
 
  \begin{table}
 	\small
 	\centering
 	
 	 	\renewcommand{\tabcolsep}{2pt} 
 	\renewcommand\arraystretch{0.0}
 	\begin{tabular}{lccc} 
 		\toprule   
 	    \textbf{Model}& \textbf{Time Complexity} & \textbf{Scholarly Graph} (s)	& \textbf{Music Graph} (s)\\ \midrule
	 	Spinglass& very slow &12548.68	& 10372.17\\ \midrule
 		FG& $O(|V|log^2|V|)$&280.60	& 272.34\\ \midrule
 		Louvain& linear&80.01	&63.02\\ \midrule
 		Walktrap&$O(|V|^2log|V|)$& 638.44	& 503.24\\ \midrule
 		Infomap&$O(|V|(|V|+|E|))$ &501.79	&425.63\\ \midrule
 		Bigclam&linear &57.01 	& 112.43\\ \midrule
 		DeepWalk& fast&720.56	& 688.32\\ \midrule
 		Node2vec& slow&	3508.44&3100.12\\ \midrule
 		gPCD&$O(\frac{2^dKP}{M})$ &\textbf{5.25} &\textbf{6.50}\\ 
 		\bottomrule
 	\end{tabular}
 	\caption{Running time analysis on gPCD and all baselines in seconds (s).}
 	\label{tab:complexity}
 	\vspace{-3em}
 \end{table}
 
 \subsubsection{Parameter Analysis}   \  
 
 We show how three parameters can affect our gPCD model performance in this section. They are the depth of the binary community tree $d$, genetic iteration number $T$ and user searching preference $\lambda$. Figure \ref{fig:tuning} show the overall impacts of all tuned parameters.

 \textbf{Depth on the Tree}: Figure \ref{fig:tuning}(a) to  Figure \ref{fig:tuning}(c) show how the depth of the binary community tree affects the model performance in accuracy and efficiency. From the figures, larger depth leads to a better personalized community detection result, while causes an exponential running time increase at the same time. Based on empirical studies, the upper bound of the depth is set to be 10 in this paper. While the depth selection may varies based on different graph sizes.
 
 \textbf{Convergence Analysis}: We observe the best chromosome updates in 40 iterations. From Figure \ref{fig:tuning}(d),   we can see the fitness score start to be stable after the 30 iterations, which means the best chromosome is no longer changed after around 30 iterations. Thus, we set $T = 30$ as the default iteration number in our approach.
 
 \textbf{Searching Preference}: In Figure \ref{fig:tuning}(e), $\lambda$ reflects the user searching preference whether he/she wants to explore new information or avoid redundant information. By selecting $\lambda$ from 0 to 1, we find  the F1-score are not very stable or have a clear correlation with $\lambda$. Based on the empirical experiments, we achieve the best performance on three models when $\lambda = 0.6$.
 
%

%% file: content/conclusionRevised.tex
\section{Conclusion}
To our best knowledge, the personalized community detection task proposed in this paper is the first attempt to address on detecting communities with different resolutions to meet with user need. To solve this task, we propose a model with an offline binary community tree construction step and an online genetic pruning step. A distributed version of our model is also deployed to accelerate running efficiency. Extensive experiments on two different datasets shows our model outperforms all baselines in terms of accuracy and efficiency. However, the current approach still partially relies on existing models such as Infomap and Node2vec. In the next step, we will design our own user-independent community detection and vertex \& user need representation models so that we can achieve a more integrated and unified model.
